\documentstyle[aps,pra,eqsecnum,amstex,lscape,prabib,preprint]{revtex}

\newcommand{\vphi}{\vec\phi} \newcommand{\ovec}{\vec0}
\newcommand{\sprod}{\mathinner{\ldotp}}

\newcommand{\modsq}[1]{\vert #1\vert^2}
\newcommand{\ep}{\epsilon}

\newcommand{\SIM}{\operatorname{SIM}}
\newcommand{\Sim}{\operatorname{sim}}

\newcommand{\arcth}{\operatorname{arc\,th}}
\newcommand{\sn}{\operatorname{sn}}
\newcommand{\cn}{\operatorname{cn}}
\newcommand{\arctanh}{\operatorname{arctanh}}

\begin{document}
\draft

\title{On the solutions of the $CP^{1}$ model in $(2+1)$ dimensions}
\author{A.~M. Grundland and P.~Winternitz}
\address{Centre de Recherches Math{\'e}matiques, Universit{\'e} de
Montr{\'e}al,
C. P. 6128, Succ.\ Centre-ville, Montr{\'e}al, (QC) H3C 3J7, Canada}
\author{W.~J. Zakrzewski}
\address{Department of Mathematical Sciences, University of Durham, Durham,
DH1 3LE, UK }
\date{\today}
\maketitle
\begin{abstract}
We use the methods of group theory to reduce the equations
of motion of the $CP^{1}$ model in (2+1) dimensions to sets of two
coupled ordinary differential equations. We
decouple and solve many of these equations in terms of
elementary functions, elliptic functions and Painlev{\'e}
transcendents. Some of the reduced equations do not have the Painlev{\'e}
property thus indicating that the model is not integrable,
while it still posesses many properties of integrable systems (such as
stable ``numerical'' solitons).
\end{abstract}

\section{Introduction}\label{sec1}

Over the past few years, it has become clear that many physical
processes can be described well in terms of various partial nonlinear
differential equations.  The areas providing such equations, range from
solid state physics, hydrodynamics and particle physics to biophysics
and biochemistry.  As the equations are nonlinear, in general, they
are hard to solve; in fact, so far no general method of solving these
equations is known and each equation has to be treated on its
own. However, a particular class of equations, which are derived from
the so-called integrable models can be solved using some very general
techniques\cite{1} (inverse scattering methods, Bdcklund
transformations,\dots). These equations are very special and their
solutions have very special properties. Many of these equations have
solutions which are localised in space and propagate at a constant
speed. Such solutions, usually called solitons, or extended structures
in general, have received a lot of attention in recent years.
However, most of these equations depend on two space (or space-time)
variables and, as such, can only describe phenomena which are
quasi-twodimensional. When they involve more variables, either all
variables come in a very nonsymmetric way or the models are very
special.

Most applications in nature involve 3 spatial dimensions, and in many
applications all spatial variables come on an equal footing. When
the applications involve, for example, particle physics or relativity,
the underlying models are Lorentz covariant. Such models are,
generally, nonintegrable and the methods mentioned above do not apply.
On the other hand, some of them can be studied numerically. Such
studies have involved full simulations of similar models in (2+1)
dimensions or simulations of various reduced models (i.e.\
approximations to the original models). Some of these
studies\cite{2} have found that even though the models were
not integrable the behaviour of their extended structures resembled
the behaviour that was expected had the models been integrable (i.e.\
the structures preserved their shapes well, there was little radiation
etc.) Morever, the approximate methods\cite{3}
gave results which were virtually indistinguishable from the results
of full simulations.  Hence we feel that the behaviour of integrable
models may not be that very unusual; other models which, strictly
speaking, are not integrable maybe be ``almost" so; with their
extended structures showing in their behaviour very little difference
from what may be expected had the models been integrable.  Moreover,
various approximate methods work well and may be used to provide some
insight to the behaviour of the solutions of the full equations.

Most of the results, which involve models in (2+1) dimensions, were obtained
in the so-called $CP^{1}$ model (also called the $S^{2}$ or $O(3)$ model)
and its modifications.\cite{2} This model, in its original
version\cite{4}, is probably the simplest model
in (2+1) dimensions which is relativistically covariant and which admits
the existence of localised soliton-like solutions.

 The $CP^{1}\,\sigma$ model is defined by the Lagrangian density
\begin{equation}
\mathcal{L}_{\sigma} =\tfrac14(\partial^\mu \vphi)
\cdot (\partial_\mu \vphi), \label{eq1.1}
\end{equation}
together with the constraint $\vphi\sprod\vphi=1$.
The Euler-Lagrange equations derived from (\ref{eq1.1}) are
\begin{equation}
\partial^\mu \partial_\mu \vphi + (\partial^\mu \vphi \cdot \partial_\mu \vphi)
\vphi = \ovec. \label{eq1.2}
\end{equation}

The model can be modified by the addition of further terms to the
Lagrangian density. The terms that have been studied the most extensively
involve the (2+1) dimensional analogue of the ``Skyrme'' term\cite{5} and
various ``potential'' terms\cite{2,6}. They were added, primarily, to
stabilise the soliton-like structures. In the original model the
soliton-like structures were not
really stable; any perturbation would induce their shrinking, or expanding
which they could do without any cost of energy due to the conformal
invariance of the pure $CP^{1}$ model. Apart from curing the shrinking, and
inducing also weak forces between soliton-like structures the additional
forces had little effect on the dynamics of these structures.
Moreover, the affects of nonintegrability of these models, were also not that
different from similar effects in the pure $CP^{1}$ case.
Hence the dynamics of these models was described  well by the dynamics
of the $CP{^1}$ case. The same was true when one looked at the
approximate methods \cite{7}.

These observations suggest that a lot can be learnt from looking
at exact solutions of the $CP^{1}$ model using the group theoretical
method of symmetry reduction \cite{8,9,10}.
This method exploits the symmetry of the original equations to find solutions
invariant under some subgroup
of the symmetry group  (the classic example one can give here involves
seeking solutions in two dimensions which are rotationally invariant).
The method puts all such attempts on a unified footing
and it has been applied with success to many equations.
 The method leads to equations whose
solutions represent specific solutions of the full equations; the solutions
are determined locally and the method does not tell us whether these solutions
are stable or not with respect to any perturbations.

In our case we would like to apply this method to looking for solutions
of the original $CP^{1}$ model; from the remarks made above we can hope
that these solutions will be also approximate solutions of the modified
models. Their stability is harder to predict; but again, guided by the
experience from the numerical simulations we hope that, at least, some of them
will be stable with respect to small perturbations.

In order to perform the symmetry reductions of the pure $CP^{1}$ model
in (2+1) dimensions we have to decide what variable to use. To avoid
having to use the constrained variables ($\vec\phi$) it is convenient
to use the $W$ formulation of the model which involved the
stereographic projection of the sphere $\vec{\phi}\cdot \vec{\phi}=1$
onto the complex plane.  In this formulation instead of using the
$\phi$ fields we express all the dependence on $\phi$ in terms of
their stereographic projection onto the complex plane $W$. The $\phi$
fields are then related to $W$ by
\begin{equation}
 \phi^1 =  {W + W^* \over 1 + \modsq{W}},\quad
 \phi^2 = i {W - W^* \over 1 + \modsq{W}},\quad
 \phi^3 =  {1 - \modsq{W} \over 1 + \modsq{W}}.
\label{eq1.3}
\end{equation}
In this formulation the Lagrangian density becomes
\begin{equation}
L={\partial_\mu  W\partial^\mu  W^*\over (1+\vert W\vert^{2})^{2}},
\label{eq1.4}
\end{equation}
where $^{\ast}$ denotes complex conjugation.

To perform our analysis it is convenient to use the polar version of
the $W$ variables; i.e.\ to put $W=R\exp(i\psi)$ and then study the
equations for $R$ and $\psi$. The advantage of this approach is that
the equations become simple; the disadvantage comes from having to pay
attention that $R$ is real and $\psi$ should be periodic with a period
of $2\pi$. (If the period is not $2\pi$ then the solution may become
multi-valued etc.)  Thus if we find solutions that do not obey these
restrictions, then these solutions, however interesting they may be,
cannot in general be treated as solutions of the original model.

The equations for $R$ and $\psi$ take the form
\begin{equation}
\partial_{tt}\psi-\partial_{xx}\psi-\partial_{yy}\psi
 +2{(1-R^2)\over R(1+R^2)}(
\partial_t \psi \partial_t R -
\partial_x \psi \partial_x R -
\partial_y \psi \partial_y R)=0
\label{eq1.5}
\end{equation}
and
\begin{multline}
\partial_{tt}R-\partial_{xx}R-\partial_{yy}R -{R(1-R^2)\over (1+R^2)}(
 \partial_t \psi \partial_t \psi - \partial_x \psi \partial_x \psi-
 \partial_y \psi \partial_y \psi)\\ -{2R\over (1+R^2)}\bigl((
 \partial_t R)^2 - (\partial_x R)^2 -(\partial_y
 R)^2\bigr)=0.
\label{eq1.6}
\end{multline}

Note, that if we put $R=1$ the second equation is automatically
satisfied and the first one reduces to
\begin{equation}
\partial_{tt}\psi-\partial_{xx}\psi-\partial_{yy}\psi=0
\label{eq1.7}
\end{equation}
i.e.\ the linear wave equation for the phase $\psi$.

In section~\ref{sec2} we determine the symmetry group of
equations~(\ref{eq1.5}) and~(\ref{eq1.6}).  In the following section
we present coupled pairs of reduced ordinary differential equations
(ODE's) for all two-dimensional subgroups of the symmetry group.
Sections~\ref{sec4} and~\ref{sec5} are devoted to the presentation of
explicit solutions.  We finish the paper with a short discussion of
the derived solutions, their relation to the solutions known
before and their physical relevance.

\section{The Symmetry Group and its Subgroups}\label{sec2}

The symmetry group of the system~(\ref{eq1.5}) and~(\ref{eq1.6}) can
be calculated using standard methods\cite{8,9,10}.  We actually made
use of a MACSYMA package\cite{11} that provides a simplified and
partially solved set of determining equations.

Solving those we find that the symmetry group has the structure of a direct
product, namely,
\begin{equation}
G\sim \SIM(2,1)\otimes SU(2),
\label{eq2.1}
\end{equation}
where $\SIM(2,1)$ is the similitude group of (2+1) dimensional
Minkowski space (the Poincar{\'e} group extended by dilations).
The group $SU(2)$ rotates the components of the fields amongst each other.

The corresponding Lie algebras $\Sim(2,1)$ and $su(2)$ can be represented
by vector fields acting on $R$ and $\psi$ and the space-time coordinates.
A suitable basis is given by two Lorentz boosts $K_1$ and $K_2$,
one rotation $L$, three translations $P_0$, $P_1$ and $P_2$, one dilation
$D$ and three $su(2)$ generators $X$, $Y$ and $Z$.
We have
\begin{gather}
\begin{gathered}
K_1=-(x\partial_t+t\partial_x),\quad
K_2=-(y\partial_t+t\partial_y),\quad
L=-x\partial_y+y\partial_x\\
P_0=\partial_t,\quad P_1=\partial_x,\quad
P_2=\partial_y,\quad D=t\partial_t+x\partial_x+y\partial_y,
\end{gathered}
\label{eq2.2}\\
\begin{gathered}
X={1\over 2}\biggl(\Bigl(R-{1\over R}\Bigr) \sin\psi\partial_{\psi}
+\bigl(R^2+1\bigr) \cos\psi\partial_R\biggr),\\
Y={1\over 2}\biggl(\Bigl(R-{1\over R}\Bigr) \cos\psi\partial_{\psi}
-\bigl(R^2+1\bigr) \sin\psi\partial_R\biggr),\\
Z = \partial_{\psi}.
\end{gathered}
\label{eq2.3}
\end{gather}

Our aim is to obtain solutions of eq.~(\ref{eq1.5}) and~(\ref{eq1.6}
by the method
of symmetry reduction\cite{8,9,10}.
 In practice we shall
require that solutions are invariant under a two-dimensional subgroup
of the symmetry group $G$. This will reduce the original partial
differential equations~(\ref{eq1.5}), (\ref{eq1.6}) to a system of ODEs.
Subalgebras of $\Sim(2,1)$ were classified in ref.~\cite{12}.
 A two-dimensional algebra $\{\hat A,\hat B\}$ can be either Abelian,
$[\hat A,\hat B]=0$, or solvable non-Abelian, $[\hat A,\hat B]=\hat A$.
Subalgebras of the direct sum $\Sim(2,1)\oplus su(2)$ can be obtained
by the Goursat ``twist'' method.\cite{13,14}

The result is the following. There exist 10 parametric classes
of Abelian subalgebras, represented by
\begin{equation}
\{\hat A+aZ,\quad \hat B+bZ\}
\label{eq2.4}
\end{equation}
with $a,b\in \Bbb R$, $Z$ as in (\ref{eq2.3}) and $\{\hat A,\hat
B\}$ equal to one
of the following pairs:
\begin{equation}
\begin{gathered}
\{K_1,P_2\};\quad \{D,K_2+L\}; \quad \{K_2+L,P_0-P_1\};\quad
\{ P_2,P_0-P_1\};\quad\{
L,P_0\};\\
\{P_1,P_2\};\quad \{P_0,P_1\};\quad\{ D,K_1\};\quad\{ D,L\};\quad
\{D-K_1,P_0-P_1\}.
\end{gathered}
\label{eq2.5}
\end{equation}
Further, there exist 15 parametric classes of nonabelian
two-dimensional subalgebras represented by
\begin{equation}
\hat A+cZ, \hat B\qquad c\in \Bbb R,
\label{eq2.6}
\end{equation}
where $\{\hat A, \hat B\}$ is one of the pairs:
\begin{equation}
\begin{gathered}
\{K_1,K_2+L\}; \{D,P_2\};\{ K_1,P_0-P_1\};\{D,P_0-P_1\};
\{K_1+\epsilon P_2,P_0-P_1\};\\
 \{D+\epsilon(K_2+L),P_0-P_1\};
\{D,P_0\};\{ D+aK_1,K_2+L\};\{ D+aK_1,P_2\};\\
\{ D-K_1+\epsilon(P_0-P_1),
K_2+L\};\{ D+K_2+\epsilon(P_0+P_2),P_1\}; \{D+aL,P_0\};\\
\{D+aK_1,P_0-P_1\}; \{D+K_1+\epsilon(P_0+P_1), P_0-P_1\};\\
\{D+{1\over 2}K_1,K_2+L+\epsilon(P_0+P_1)\}
\end{gathered}
\label{eq2.7}
\end{equation}
with $a\in \Bbb R$ and $\epsilon=\pm 1$.

\section{The Reduced Equations}\label{sec3}

For each subgroup (\ref{eq2.4}) and (\ref{eq2.6}) we find three invariants,
$\xi$, $R$
and $F$, using standard methods.\cite{8,9,10} In
terms of these we express the two functions $R$ and $\psi$ of (\ref{eq1.5})
and (\ref{eq1.6}) as
\begin{equation}
R=R(\xi),\quad \psi=\alpha(x,y,t)+F(\xi),\quad \xi=\xi(x,y,t),
\label{eq3.1}
\end{equation}
where $\alpha$ and $\xi$ are given for each subalgebra in
Table~\ref{tab1} and~\ref{tab2}. Derivatives with respect to the variable $\xi$
will be denoted by dots.
We introduce the two invariant operators $\Delta$ and $\nabla^2$ by
setting
\begin{equation}
\Delta f=f_{tt}-f_{xx}-f_{yy},\quad (\nabla
f,\nabla g)=f_tg_t-f_xg_x-f_yg_y
\label{eq3.2}
\end{equation}
and consider three cases separately.
\begin{enumerate}
\item $(\nabla \xi)^2\ne0$.
We put
\begin{equation}
{\Delta \xi\over (\nabla \xi)^2}=-p=-{\dot g\over g},\quad {(\nabla
  \alpha ,\nabla \xi)\over (\nabla
\xi)^2}=h,\quad {\Delta \alpha\over (\nabla \xi)^2}=s,\quad {(\nabla \alpha)
^2\over (\nabla \xi)^2}=l,
\label{eq3.3}
\end{equation}
with $g=g(\xi)$, $h=h(\xi)$, $s=s(\xi)$ and $l=l(\xi)$.

For further use we also introduce
\begin{equation}
m=\dot h-{\dot g\over g}h-s\label{eq3.4}
\end{equation}
and
\begin{equation}
B=2{h^2-l\over g^2}.
\label{eq3.5}
\end{equation}

The two PDE's (\ref{eq1.5}) and (\ref{eq1.6}) now reduce to
\begin{gather}
{\ddot F+2\dot R\dot F{(1-R^2)\over R(1+R^2)}
-{\dot g\over g}\dot F+2{(1-R^2)\over
 R(1+R^2)}\dot R h +s=0}
\label{eq3.6}\\
\ddot R-{2R\over 1+R^2}\dot R^2-{R(1-R^2)\over 1+R^2}\dot F^2
-{\dot g\over g}\dot R -2{R(1-R^2)\over 1+R^2}\dot F h -{R(1-R^2)\over
 1+R^2}l=0\label{eq3.7}
\end{gather}
For each algebra the
 functions (\ref{eq3.3}) are given in Table~\ref{tab1}.

In order to solve the above system  we must decouple its two equations.
Putting $\dot F+h=V$ we first rewrite equation~(\ref{eq3.6}) as
\begin{equation}
\dot V+2\dot R{(1-R^2)\over R(1+R^2)}V-{\dot g\over g}V-m=0
\label{eq3.8}
\end{equation}
with $m$ as in eq.~(\ref{eq3.4}).

For $m=0$ we solve (\ref{eq3.6}) and obtain
\begin{equation}
\dot F\,=A{(1+R^2)^2\over R^2}g(\xi)-h,\qquad  A=\text{const}.
\label{eq3.9}
\end{equation}

Next we substitute $\dot F$ into  equation~(\ref{eq3.7})
and obtain a second order ODE for $R(\xi)$
\begin{equation}
\ddot R-{2R\over 1+R^2}\dot R^2-{\dot g\over g}\dot R
  -A^2g^2{(1-R^2)(1+R^2)^3\over R^3}
+{R(1-R^2)\over 1+R^2}(h^2-l)=0.
\label{eq3.10}
\end{equation}

For $m\ne0$ eq.~(\ref{eq3.8}) is inhomogeneous.
We can still decouple it
by putting
\begin{gather}
\dot F = {Um\over \dot U} - h
\label{eq3.11}\\
\dot U = {mR^2\over (R^2+1)^2 g}.
\label{eq3.12}
\end{gather}

Using eq.~(\ref{eq3.11}) we can rewrite (\ref{eq3.7})
as a third order ODE for the auxiliary function $U(\xi)$. If we can solve
it we obtain $R(\xi)$ from (\ref{eq3.12}).
However, in this paper we restrict our attention to the case of $m(\xi)=0$.

\item $(\nabla\xi)^2=0$, $\Delta\xi=0$, $(\nabla\alpha, \nabla \xi)\ne0$.
The reduced equations decouple immediately and we have
\begin{equation}
\begin{gathered}
(\nabla \alpha, \nabla \xi) {(1-R^2)\over R(1+R^2)}\dot R
        = -{1 \over 2} \Delta \alpha,\\
(\nabla \alpha, \nabla \xi)
\dot F = -{1\over 2}(\nabla\alpha)^2.
\end{gathered}
\label{eq3.17}
\end{equation}
For $(\nabla\alpha, \nabla\xi) \ne 0$ we can integrate directly to obtain
\begin{gather}
{R\over R^2+1}=C\exp\biggl\{-{1\over 2}\int{\Delta\alpha\over
(\nabla \alpha ,\nabla \xi)}\,d\xi\biggr\}
\label{eq3.18}\\
F=-{1\over 2}\int {(\nabla \alpha)^2\over (\nabla \alpha,\nabla
  \xi)}\,d\xi + F_0.
\label{eq3.19}
\end{gather}
For $(\nabla\alpha,\nabla\xi)=0$ we must have also $\Delta\alpha=0$,
$(\nabla\xi)^2=0$. Then $R(\xi)$ and $F(\xi)$ are arbitrary functions
of $\xi$. In particular this is true for $\alpha=0$.

\item $(\nabla \xi)^2=0$, $\Delta \xi = h(\xi) \ne 0$.  This case can actually
not occur, since these two equations are compatible only for $h(\xi) = 0$.

\end{enumerate}

\section{Analysis of Second Order ODE}\label{sec4}

\subsection{General Comments}

In order to obtain explicit solutions we need to solve the ODE (\ref{eq3.10})
for the function $R(\xi)$. This equation is in the class analysed by
Painlev{\'e} and Gambier\cite{15,16,17}; namely
it is of the form
\begin{equation}
\ddot y=f(\dot y,y,x)
\label{eq4.1}
\end{equation}
where $f$ is rational in $\dot y$ and $y$ and analytical in $x$.
If this equation has the Painlev{\'e}  property (no movable singularities
other than poles) then it can be transformed into one of the 50 standard
equations listed e.g.\ by Ince\cite{17}. The Painlev{\'e}
test\cite{18,19}
provides us with  necessary (but not sufficient) conditions for
eq.~(\ref{eq4.1}) to have the Painlev{\'e}  property. The solution is expanded
about an arbitrary
point $x_0$ in the complex $x$-plane in a Laurent series
\begin{equation}
y(x)=\sum_{k=0}^{\infty} a_k \tau^{k+p},\quad \tau=x-x_0,
\label{eq4.2}
\end{equation}
where $p$ is required to be an integer (usually a negative one). The
coefficients
are obtained from a recursion relation of the form
\begin{equation}
P_k a_k=h_k(x_0,a_0,a_1,\dots a_{k-1}).
\label{eq4.3}
\end{equation}

Since (\ref{eq4.2}) is supposed to represent a general solution it must depend
on two constants; $x_0$ and $a_r$ for some nonnegative $r$, called a resonance
value. This occurs if the function $P_r$ satifies $P_r=0$. Then $a_r$ is
arbitrary and we have a consistency condition, the ``resonance condition",
$h_r(x_0,a_0,a_1,\dots a_{r-1})=0$ (which must be satisfied
identically in $x_0$).
If the above conditions are satisfied, the Painlev{\'e} test is passed
and (\ref{eq4.2}) represents a two parameter family of formal solutions
(locally, within the radius of convergence of the series).

Turning our attention to eq.~(\ref{eq3.10}) we note that the cases
$A\ne 0$ and
$A=0$ must be treated  separately.

\subsection{The case $A\ne0$}

The Painlev{\'e}  test applied directly fails immediately since a balance
between the most singular terms occurs for $p=-\frac{1}{2}$, i.e.\ we have
a movable square root branch point. To remedy this problem we put
\begin{equation}
R(\xi)=\sqrt{-U(\xi)},
\label{eq4.4}
\end{equation}
and obtain
\begin{equation}
\ddot U=\dot U^2\Bigl[{1\over 2U}+{1\over U-1}
\Bigr]+{2A^2}g^2{(1+U)(1-U)^3\over U}
+\dot U{\dot g\over g}+2(h^2-l)
{U(1+U)\over (U-1)}.
\label{eq4.5}
\end{equation}

We can now choose a new variable $\eta$ to be
\begin{equation}
\eta=\int g(\xi)\,d\xi
\label{eq4.6}
\end{equation}
and  transform eq.~(\ref{eq4.5}) into
\begin{equation}
\ddot U=
\dot U^2\Bigl[{1\over 2U}+{1\over U-1}
\Bigr]+2{A^2} {(1+U)(1-U)^3\over U}
+B{U(1+U)\over (U-1)}
\label{eq4.7}
\end{equation}
with $B$ as in (\ref{eq3.5}).

For $B$=const this is eq.~PXXXVIII listed e.g.\ by Ince\cite{17}.
It has a first integral  $K$
that we use to write a first order equation
for $U$:
\begin{equation}
\dot U^2=-4A^2U^4+4KU^3+(8A^2-2B-8K)U^2+4KU-4A^2.
\label{eq4.9}
\end{equation}
Since we have $A\ne0$ we can rewrite (\ref{eq4.9}) as
\begin{equation}
(\dot U)^2=-4A^2(U-U_1)(U-U_2)(U-U_3)(U-U_4),
\label{eq4.10}
\end{equation}
where the constant roots of the right hand side of
(\ref{eq4.10}) $U_i$, $i=1\dotsc 4$ satisfy
\begin{equation}
\begin{gathered}
U_1U_2U_3U_4= 1\\
U_1U_2U_3+U_1U_2U_4+U_1U_3U_4+U_2U_3U_4 = {K\over A^2}\\
U_1U_2+U_1U_3+U_1U_4+U_2U_3+U_2U_4+U_3U_4= -2
+{B\over 2A^2}+{2K \over A^2}\\
U_1+U_2+U_3+U_4={K\over A^2}.
\end{gathered}
\label{eq4.11}
\end{equation}

Eq.~(\ref{eq4.10}) has elementary algebraic and trigonometric
solutions, as well as solutions
which resemble solitary waves or kink-like structures
(in the symmetry variable $\eta$) in the case
of multiple roots. If all $U_i$ are distinct then the solutions
of (\ref{eq4.9}) involve elliptic functions \cite{20}. Explicit solutions will
be presented in the next section.

If $B$ in eq.~(\ref{eq4.9}) is not constant we proceed differently.
We again introduce a new independent variable
\begin{equation}
\eta=e^{\int g\,d\xi}
\label{eq4.12}
\end{equation}
and transform eq.~(\ref{eq4.5}) into
\begin{equation}
\ddot U=\dot U^2\Bigl[{1\over 2U}+{1\over U-1}
\Bigr]-{\dot U\over \eta}+{{2A^2\over \eta^2}}(1-U)^2({1\over U}-U)
+2{h^2-l\over g^2\eta^2}{U(U+1)\over U-1}.
\label{eq4.13}
\end{equation}
For
\begin{equation}
2{h^2-l\over g^2\eta^2}=\delta=\text{const}
\label{eq4.14}
\end{equation}
this is the equation for the fifth Painlev{\'e} transcendent.

The values $m=0$ and $B\ne\text{const}$ occur for the cases 1 and 2 from
Table~\ref{tab1} and we actually have
\begin{equation}
\eta=\xi,\qquad \delta=2\Bigl(b^2\pm {a^2\over \eta^2}\Bigr),
\label{eq4.15}
\end{equation}
so $\delta=$const requires $a=0$.

Hence we obtain the solutions
\begin{equation}
\begin{gathered}
U(\xi)=P_V(\alpha,\beta,\gamma,\delta;\xi)\\
\alpha=-\beta=2A^2,\quad \gamma=0,\quad \delta=2b^2,
\end{gathered}
\label{eq4.16}
\end{equation}
for equations describing the group reductions 1 and 2 (of
Table~\ref{tab1}) in the
case when $a=0$.

For $b=0$ in the cases of these two reductions we have $B=2a^2=\text{const}$
and we obtain solutions in terms of equation~(\ref{eq4.9}).

\subsection{The case $A=0$, $B=\text{const}$}

The transformation~(\ref{eq4.4}) can also be performed for the case $A=0$.
We use the first integral to write our equation as
\begin{equation}
\begin{gathered}
\dot U^2=4K(U-U_1)(U-U_2)\\
U_{1,2}=L+1\pm \sqrt{L(L+2)},\quad L={B\over 4K},\quad K\ne0.
\end{gathered}
\label{eq4.17}
\end{equation}

For $K=0$ we find a solution immediately; namely we have
\begin{equation}
U=U_0 e^{\pm \sqrt{-2B}\eta},\quad B\le0.
\label{eq4.18}
\end{equation}

On the other hand, for $A=0$ in eq.~(\ref{eq3.10}), the Painlev{\'e} expansion
(\ref{eq4.2}) gives us $p=-1$ for the leading (most singular) power. Hence
the transformation (\ref{eq4.4}) is not required and so we can transform
eq.~(\ref{eq4.2}) directly into one of the standard forms.

We put
\begin{equation}
R=-i{Z(\eta)+\mu (\xi)\over Z(\eta)-\mu (\xi)},\quad
\eta=\eta(\xi),\quad \mu \ne0,\quad \dot\eta\ne0
\label{eq4.19}
\end{equation}
and obtain
\begin{equation}
\ddot Z={\dot Z^2\over Z}-{1\over \dot \eta}\bigl({\ddot \eta\over \dot \eta}
-{\dot g\over g}\bigr)\dot Z+{1\over \dot \eta^2\mu }\bigl(\ddot \mu
-{\dot \mu ^2\over \mu }-{\dot g\over g}\dot \mu \bigr)Z+{h^2-l\over 4\dot
\eta^2\mu ^2}{Z^4-\mu ^4\over Z}.
\label{eq4.20}
\end{equation}

If $B$ of eq.~(\ref{eq3.5}) is constant we can choose $\eta$ as in
eq.~(\ref{eq4.6}), set
$\mu =1$ and obtain the equation
\begin{equation}
\ddot Z={\dot Z^2\over Z}+{B\over 8}\Bigl({Z^4-1\over Z}\Bigr).
\label{eq4.21}
\end{equation}
This equation can be integrated directly for $B=0$. For $B\ne0$ it has a first
integral $K$ in terms of which we obtain  a first order ODE
\begin{equation}
\begin{gathered}
\dot Z^2={B\over 8}(Z^2-Z_1^2)(Z^2-Z_2^2)\\
Z_{1,2}^2=-K\pm \sqrt{K^2-1}.
\end{gathered}
\label{eq4.22}
\end{equation}
We again have elliptic function solutions. They are however not new.
Since $R$ is real (and nonnegative) we must require that
\begin{equation}
Z=e^{i\sigma(\eta)}, \quad \text{for } \mu =1, 0\le\eta<2\pi.
\label{eq4.23}
\end{equation}
The relation between the function $\sigma(\eta)$ and $U(\eta)$ of
eq.~(\ref{eq4.1}) is
\begin{equation}
\sqrt{-U(\eta)}=\Bigl\vert {\sin\sigma\over 1-\sin\sigma}\Bigr\vert.
\label{eq4.24}
\end{equation}

For $B\ne$const we use the variable~(\ref{eq4.11}), again set $\mu =1$ and
reduce
eq.~(\ref{eq4.20}) to
\begin{equation}
\ddot Z={\dot Z^2\over Z}-{1\over \eta}\dot Z+{h^2-l\over
4g^2\eta^2}{Z^4-1\over Z}.
\label{eq4.25}
\end{equation}

For $\delta$, defined by~(\ref{eq4.14}), being constant ($\delta=\delta_0$)
 eq.~(\ref{eq4.25})
is a special case of the third Painlev{\'e} transcendent
$P_{\text{III}}(\alpha,\gamma,\beta, \delta:\eta)$ and so we have as a solution
of eq.~(\ref{eq4.25})
\begin{equation}
Z=P_{\text{III}}\Bigl(0,0,{\delta_0\over 8},-{\delta_0\over
8};\eta\Bigr).
\label{eq4.26}
\end{equation}
This equation is, however, not new; it can be transformed into
solution (\ref{eq4.16}) by making use of
relations between special cases of $P_{\text{V}}$ and
$P_{\text{III}}$.

\subsection{Comments on the Painlev{\'e} analysis and integrability of model}

For $A\ne0$ eq.~(\ref{eq4.5}) passes the Painlev{\'e} test for any
function $g(\xi)$ and $h^2(\xi)-l(\xi)$. Indeed, we find $p=-1$ in the
expansion~(\ref{eq4.2}).
A resonance is obtained for $k=1$; the resonance condition is satisfied
and so the coefficient $a_1$ is a free constant (as is $x_0$).

The Painlev{\'e} test only checks whether certain necessary conditions are
satisfied. If an equation of the type~(\ref{eq4.1}) does actually have the
Painlev{\'e} property (as opposed to merely passing the Painlev{\'e} test),
then it can be transformed into a standard form by a M{\"o}bius transformation
(with variable coefficients)
\begin{equation}
y(\xi)={\alpha(\xi)U(\eta(\xi))+\beta(\xi)\over
\delta(\xi)U(\eta(\xi))+\rho(\xi)},\quad \eta=\eta(\xi),\quad
\alpha\rho-\beta\delta=\pm 1.
\label{eq4.27}
\end{equation}

Eq.~(\ref{eq4.5}) is already, to a large extend, standardized. Indeed,
the coefficient
of $\dot U^2$ has poles at $U=0$, 1 and $\infty$. This puts the equation into
Ince's class~IV and the residues have the correct
values. Hence we have
$\alpha=\rho=1$, $\beta=\delta=0$ in eq.~(\ref{eq4.27}). The only remaining
permitted transformation is that of the independent variable. We have shown
above that eq.~(\ref{eq4.5}) can be reduced to the elliptic equation
if and only if $B$ in eq.~(\ref{eq4.9}) is constant. It can be
transformed into
the equation for the $P_{\text{V}}$ transcendent if and only
if its coefficients satisfy
\begin{equation}
{d\over d\xi}\bigl({h^2-l\over g^2}\bigr)-2\bigl({h^2-l\over g^2}\bigr)g=0.
\label{eq4.28}
\end{equation}
In all other cases the equation~(\ref{eq4.5}) cannot be transformed into
a standard form and hence it does not have Painlev{\'e} property.

The situation is exactly the same for $A=0$. Eq.~(\ref{eq3.10}) passes the
Painlev{\'e} test and is transformed into eq.~(\ref{eq4.20}) by a M{\"o}bius
 transformation. The coefficient of $\dot Z^2$ has poles at
$Z=0$ and $Z\rightarrow\infty$ with the correct residues. Hence only
$Z(\xi)\rightarrow \alpha(\xi)Z\bigl(\eta(\xi)\bigr)$ is
permitted. Eq.~(\ref{eq4.20}) is of Ince's
type~II and
has the Painlev{\'e} properties if and only if eq.~(\ref{eq4.28}) is satisfied.

Thus we have shown that eq.~(\ref{eq4.5}) has the Painlev{\'e} property
if and only if the coefficients satisfy
$B=\text{const}$, or Eq.~(\ref{eq4.28}).

\section{Explicit Solutions}\label{sec5}

We have reduced the original system (\ref{eq1.5}), (\ref{eq1.6}) for
the function
\begin{equation}
  \label{eq5.1a}
  W=Re^{i\psi}
\end{equation}
to one of the   pairs of equations \{ (\ref{eq3.9}),
(\ref{eq3.10})\}, or $\{(3.14), (3.15)\}$.

Let us first look at the pair \{(\ref{eq3.9}),
(\ref{eq3.10})\}. Eq.~(3.9) provides $F(\xi)$ by a quadrature,
once eq.~(\ref{eq3.10}) is solved. In section~\ref{sec4} we have
further reduced eq.~(\ref{eq3.10}). Using eq.~(\ref{eq4.4}) we replace
equations for $R(\xi)$ by equations for $U(\xi)$, where $U(\xi)$ must satisfy
$U(\xi)\le 0$.

As mentioned above, the algebras~1 and~2 of Table~\ref{tab1} lead to
solutions of the form~(\ref{eq4.16}), i.e.\ the fifth Painlev{\'e}
transcendent $P_{\text{V}}(\xi)$, for $a=0$, $b\ne 0$.

Algebras 1--23 lead to the elliptic function equation~(4.9)
for $m=0$, $A\ne 0$, $B=\text{const}$ and to eq.~(\ref{eq4.17}) for
$m=0$, $A=0$, $B=\text{const}$, $K\ne 0$. In both cases, elementary
solutions are obtained in the case of multiple roots of the polynomial
on the right hand side of the equation. Let us discuss in more detail
the real nonsingular solutions of these equations. The character of
the solution depends crucially on the sign of $B$ in the table ($B$
is defined in (\ref{eq3.5})). We have
\begin{itemize}
\item $B > 0$ for algebras
  \begin{equation}
    \label{eq5.1}
\begin{gathered}
    1\ (b=0,a\ne 0), 10\ (a^2+b^2\ne 0), 11\ (a\ne 0), 12\ (a\ne 0),\\
15\ (a^2>b^2), 16\ (b\ne 0), \\
17\ (a=0, b\ne 0, x^2 + y^2 - t^2 > 0), 18\ (a = 0, b\ne 0, x^2 + y^2 - t^2 >
0), \\ 19\ (a = 0, b\ne 0, x^2 + y^2 - t^2 > 0).
\end{gathered}
  \end{equation}
\item $B<0$ for algebras
  \begin{equation}
    \label{eq5.2}
\begin{gathered}
    2\ (b=0, a\ne 0), 13\ (a\ne 0), 14\ (ab \ne 0),\\
 15\ (a^2-b^2<0),  \\
17\ (a=0, b\ne 0, t^2 - x^2 - y^2 > 0), 18\ (a = 0, b\ne 0, t^2 - x^2 - y^2  >
0), \\ 19\ (a = 0, b\ne 0, t^2 - x^2 - y^2 > 0).
\end{gathered}
  \end{equation}
\end{itemize}
In all other cases with $m=0$ we have $B=0$

Let us first run through all elementary functions solutions,
remembering that the independent variable is $\eta$ given in
eq.~(\ref{eq4.6}).

Localized solutions are obtained precisely for the algebras
(\ref{eq5.1}). From eq.~(\ref{eq4.17}) (i.e.\ $A=0$) we obtain a kink
in $R(\xi)$ where $\xi$ and function $h(\xi)$ are read off from
Table~\ref{tab1}. Two cases are to be considered.

\begin{enumerate}
\item[I.] $A=0$, $L=-2$, $K<0$, $B>0$:
\end{enumerate}

The solution is:
\begin{enumerate}
\item
\begin{equation}
\label{eq5.3}
R=\pm \tanh \frac{1}{2}\sqrt{\frac{B}{2}}(\eta-\eta_0), \quad F=-\int
h(\eta)\,d\eta +F_0
\end{equation}
\end{enumerate}

\begin{enumerate}
\item[II.] $A\ne 0$, $B> 4(A^2-K) >0$, $K<0$:
\end{enumerate}

Eq.~(\ref{eq4.10}) ($A\ne 0$) leads to solitary wave (``bump'' or
``well'' type solutions) for the function $U(\eta)$ in the following cases:
\begin{enumerate}
\setcounter{enumi}{1}
\item $U_4=U_3=U_2<U\le U_1<0$
\begin{equation}
\label{eq5.5}
U=U_2+\frac{U_1-U_2}{1+(U_1-U_4)^2A^2(\eta-\eta_0)^2}.
\end{equation}
Eq.~(\ref{eq5.5}) represents an ``algebraic bump''.
\item $U_4\le U<U_3=U_2=U_1<0$
\begin{equation}
\label{eq5.6}
U=U_1-\frac{U_1-U_4}{1+(U_1-U_4)^2A^2(\eta-\eta_0)^2}.
\end{equation}
This is an ``algebraic well''.
\item $U_4<U_3=U_2<U\le U_1<0$
\begin{equation}
\label{eq5.7}
U=U_2+\frac{(U_1-U_2)(U_2-U_4)}{(U_1-U_4)\cosh^2
  A\sqrt{(U_1-U_2)(U_2-U_4)}(\eta-\eta_0)-(U_1-U_2)}.
\end{equation}
Eq.(\ref{eq5.7}) is an ``exponential bump''.
\item $U_4\le U\le U_3=U_2<U_1<0$
\begin{equation}
\label{eq5.8}
U=U_2-\frac{(U_1-U_2)(U_2-U_4)}{(U_1-U_4)\cosh^2
  A\sqrt{(U_1-U_2)(U_2-U_4)}(\eta-\eta_0)-(U_2-U_2)}.
\end{equation}
This is an ``exponential well''.
\end{enumerate}
Further elementary solutions of eq.~(\ref{eq4.10}) are
trigonometrically periodic.
\begin{enumerate}
\setcounter{enumi}{5}
\item $U_4=U_3<U_2\le U\le U_1<0$, $K<0$, $B>4(A^2-K)>0$
\begin{equation}
U=U_3+\frac{(U_1-U_3)(U_2-U_3)}{U_2-U_3+(U_1-U_2)\cos^2
  A\sqrt{(U_1-U_3)(U_2-U_3)}(\eta-\eta_0)}
\label{eq5.9}
\end{equation}
This type of solution also occurs only for the algebras~(\ref{eq5.1})
\item  $U_4 \le U \le U_3 < U_2 = U_1$
\begin{equation}
U=U_1 - \frac{(U_1-U_4)(U_1-U_3)}{U_1-U_3+(U_3-U_4)\cos^2
  A\sqrt{(U_1-U_4)(U_1-U_3)}(\eta-\eta_0)}
\label{eq5.10}
\end{equation}
This solution can occur in the case of algebras~(\ref{eq5.1}) for
$U_1<0$, i.e.\ all roots negative (and then conditions~(5.5)
hold). It can also occur for $U_3<0<U_2=U_1$ and this allows us to
have $B\le 0$, Thus, solutions~(\ref{eq5.10}) can occur for all
algebras (and variables $\xi$) 1--23 in Table~\ref{tab1}.  Notice however, that
they are periodic, rather than localized, in the variable $\eta$.
\end{enumerate}

The remaining solutions are periodic and expressed in terms of Jacobi
elliptic functions. We have:
\begin{enumerate}
\setcounter{enumi}{7}
\item $A=0$, $K>0$, $B<-8K<0$, $U_2\le U\le U_1<0$
\begin{equation}
U=\frac{U_1U_2}{U_2+(U_1-U_2)\sn^2
\left(\sqrt{-\frac{U_2K}{2}(\eta-\eta_0),k}\right)},\quad
k^2=\frac{U_1-U_2}{(-U_2)}
\label{eq5.11}
\end{equation}
This occurs for the algebras~(\ref{eq5.2})
\item $A=0$, $K<0$, $B>-8K>0$, $U_2 < U_1\le U \le 0$
\begin{equation}
R= \sqrt{-U_1}\sn\sqrt{\frac{U_2K}{2}}(\eta-\eta_0,k),\quad
k^2=\frac{U_1}{U_2}
\label{eq5.12}
\end{equation}
The algebras concerned are those of eq.~(\ref{eq5.1})
\item $A\ne 0$, $U_4\le U \le U_3 < U_2 < U_1$
\begin{equation}
\begin{gathered}
U=\frac{U_1(U_3-U_4)\sn^2[\beta(\eta-\eta_0),k]+U_4(U_1-U_3)}
{(U_3-U_4)\sn^2[\beta(\eta-\eta_0),k]+U_1-U_3},\\
k^2=\frac{(U_1-U_2)}{(U_1-U_3)}\frac{(U_3-U_4)}{(U_2-U_3)}\quad
\beta=A\sqrt{(U_1-U_3)(U_2-U_4)}
\end{gathered}
\label{eq5.13}
\end{equation}
This can occur for $U_1<0$, then we must have $B> 4(A^2+(-K))>0$,
i.e.\ the algebras~(\ref{eq5.1}). It can also occur for
$U_3<U_4<0<U_2<U_1$, then all of the algebras~1--23 of
Table~\ref{tab1} can occur.
\item $A\ne0$, $U_4<U_3<U_2\le U\le  U_1<0$
\begin{equation}
U=\frac{U_4(U_1-U_2)\sn^2[\beta(\eta-\eta_0),k]
  +U_1(U_2-U_4)}{(U_1-U_2)\sn^2[\beta(\eta-\eta_0),k]+ U_2-U_4},
\label{eq5.14}
\end{equation}
with $k^2$ and $\beta$ as in eq.~(\ref{eq5.13}). we must
have $B>4(A^2+(-K))>0$ and hence algebras~(\ref{eq5.1}).
 \item $A\ne 0$, $U_4< U< U_1$, $U_{2,3}=p\pm iq$, $q>0$
\begin{equation}
\begin{gathered}
U=\frac{[CU_4-DU_1]\cn[\beta(\eta-\eta_0),k] +
  DU_1+CU_4}{(C-D)\cn[\beta(\eta-\eta_0), k]+C+D},\\
C=(U_1-p)^2+q^2,\quad D=(U_4-p)^2+q^2\\
k^2=\frac{(U_1-U_4)^2-(C-D)^2}{4CD},\quad \beta=2A(CD)^{1/4}
\end{gathered}
\label{eq5.16}
\end{equation}

This situation can occur for all algebras 1--23 of Table~\ref{tab1}.
\item $A=0$, $K=0$, $B<0$. We obtain the solution~(\ref{eq4.18}) for
  algebras~(\ref{eq5.2}) with $\xi$ as given in Table~\ref{tab1} (and
  $\eta$ given by eq.~(\ref{eq4.6})).
\end{enumerate}

The algebras No.~24--29 of Table~\ref{tab2} correspond to
variables $\xi$ such that $(\nabla \xi)^2=0$ and hence to first order
ODEs. The solutions are readily obtained and we just list them.
\begin{xalignat}{2}
&\text{No.~24:}& R&=R_0, \psi =ay-bx+\frac{a^2+b^2}{2b}(x+t)+\psi_0,
b\ne 0 \label{eq5.17a}\\
&\text{No.~25:}& R&=c\sqrt{x+t}\pm \sqrt{c^2(x+t)-1},
\psi=\frac{b}{2}(t-x) -\frac{(a+by)^2}{2b(x+t)} +\psi_0\label{eq5.17}\\
&\text{No.~26:}& R&=c\sqrt{x+t}\pm \sqrt{c^2(x+t)-1}, \psi=b\ln
\sqrt{\frac{t^2-x^2-y^2}{x+t}} +\psi_0\label{eq5.18}\\
&\text{No.~27:}& R&=R_0, \psi=\frac{b}{2}\ln(x+t)+\psi_0\label{eq5.19}
\end{xalignat}
No.~28 and~29 provide nonconstant solutions only for $b=0$. Then
$F(\xi)$ and $R(\xi)$ are arbitrary functions of $\xi=x+t$.

All solutions presented so far are group invariant solutions in the
standard sense of the words \cite{8,9,10}.

Let us mention that the PDEs~(\ref{eq1.5}) and~(\ref{eq1.6}) can be
reduced to ODEs of the form~(\ref{eq3.6}) and~(\ref{eq3.7}), by the
transformation~(\ref{eq3.1}) where $\xi$ and $\alpha$ are any
functions satisfying eq.~(\ref{eq3.3}). The restriction is that $p$
$h$, $s$ and $l$ must be functions of $\xi$. Group theory generates
solutions of these equations by the requirement that $F$ and $\xi$ in
(\ref{eq3.1}) be invariants of subgroups of the symmetry
group. However, other solutions may exist, corresponding e.g.\ to so
called ``null variables'' \cite{21,22}, to ``conditional symmetries''
\cite{23,24}, or simply generated by the ``direct method'' of Clarkson
and Kruskal \cite{25}.

Let us just give some examples of such variables.

First a few words about null variables and the corresponding
solutions. Consider a variable $\xi$ satisfying
\begin{equation}
(\nabla \xi)^2=\Delta \xi=0
\label{eq5.20}
\end{equation}
The equations for $F(\xi)$ and $R(\xi)$ reduce to
eq.~(3.13). As mentioned in Section~\ref{sec3}, if we have also
\begin{equation}
(\nabla \alpha,\nabla \xi)=(\nabla \alpha)^2=\Delta\alpha=0
\label{eq5.21}
\end{equation}
(e.g.\ for $\alpha=\text{const}$), then $F(\xi)$ and $R(\xi)$ are
arbitrary functions. We have already encountered this situation for
$\xi=x+t$, however eq.~(\ref{eq5.20}) have more general solutions
\cite{21,22}, that can be written in terms of Riemann invariants.

Indeed let us put
\begin{equation}
\xi=H(\sigma),\quad \sigma=(\vec a, \vec x)=a_0t-a_1x-a_2y, \quad
(\vec a,\vec a)=0
\label{eq5.22}
\end{equation}
where $\vec a$ is a lightlike vector, depending on $\xi$ (i.e.\
eq.~(\ref{eq5.22}) defines $\xi$ implicitly). It is easy to check that
$\xi$ of eq.~(\ref{eq5.22}) satisfies eq.~(eq5.20) for any choice of
the function $H$ and lightlike vector $\vec a(\xi)$. The function
$H(\sigma)$ can be chosen to be $H(\sigma)=\sigma$ with no loss of
generality, since $F(\xi)$ and $R(\xi)$ are themselves arbitrary. Thus
we can replace eq.~(\ref{eq5.22}) by
\begin{equation}
\xi=a_0(\xi)t-a_1(\xi)x-a_2(\xi)y,\quad \vec a^2=0
\label{eq5.23}
\end{equation}
Choosing $\vec a$ to be constant, we recover the variable $\xi= x+t$
(up to a Lorentz transformation). Other choices give different
results, that become explicit if we can solve the algebraic
equation~(\ref{eq5.23}). For example choose
\begin{equation}
\vec a=(1,\xi,\sqrt{1-\xi^2}).
\label{eq5.24}
\end{equation}
Solving eq.~(\ref{eq5.23}) for $\xi$, we obtain
\begin{equation}
\xi=\frac{(1+x)t\pm \sqrt{(1+x)^2+y^2-t^2}}{(1+x)^2+y^2}.
\label{eq5.25}
\end{equation}

Choosing $F(\xi)$ and $R(\xi)$ appropriately, e.g.\ $F(\xi)$ constant
and $R(\xi)$ with compact support we obtain a localized solution
(localized in the variable $\xi$).

An example of a ``conditionally invariant'' solution is obtained by
putting
\begin{equation}
\psi=\psi(\xi),\quad R=R(\xi),\quad \xi=\sqrt{\frac{x^2+y^2}{x^2-t^2}}.
\label{eq5.26}
\end{equation}
We have $\alpha(x,y,t)=0$ and
\begin{equation}
p=\frac{\dot g}{g}= -\frac{1}{\xi}, \quad g=\frac{1}{\xi}, \quad
h=s=l=m=0,\quad B=0
\label{eq5.27}
\end{equation}
in eq.~(\ref{eq3.6}) and (\ref{eq3.7}). These values could hence be
added to those in Table~\ref{tab1}.

\section{Comments and Conclusions} \label{sec6}
Inserting the variables $\xi$ and $\alpha$ of Table~\ref{tab1} into the
formulas of Section~V we obtain a great variety of exact analytic solutions.

Some of our solutions are (possibly upto phase factors, contained in the
variable $\alpha$) really solutions of the $1+1$, or $2+0$ dimensional $CP^1$
model.  Thus, algebras 1, 11, 12, 16 provide solutions depending essentially
only on $x$ and $t$.  Similarly, algebras 2, 13, 14 provide essentially static
solutions (independent of $t$).  A sizable literature exists on static
solutions  [26--30].  Particularly interesting solutions of this type are
obtained for algebra 2 when we have $\xi = \sqrt{x^2 + y^2}$ and we take $b=0$.
 We obtain elliptic function solutions \eqref{eq5.11} and (5.16) as well as the
elementary solution (4.17), or more explicitly
\begin{equation} \label{6.1a}
W = W_0 \bigl( \sqrt{x^2+y^2}\bigr)^n e^{i\,n\phi}
\end{equation}
where $a=n$ is an integer (and $\phi$ is the azimuthal angle in the $xy$
plane).  This can be identified as an ``$n$-soliton solution'' (or instanton)
and it has finite energy \cite{4}.  Our static solutions are not new: they are
to be  found e.g. among those obtained by Purkait and Ray, or earlier [26--30].

The same algebra gives rise to a very different type of solution.  If we take
$b \ne 0$, $a = 0$ we express $R(\xi)$ in terms of the Painlev\'e transcendent
$P_V$, as in eq.~(4.15).  The phase is $\psi = bt + \psi_0$ so that we have a
global rotation of spins in the horizontal plane.  To our knowledge, this type
of solution is new.

Algebra 14 introduces a ``helical'' type variable $\xi$.  Solution (4.17) in
this case is
\begin{equation} \label{6.2b}
W = R_0 (x^2 + y^2)^{ab/2(1+a^2)} e^{-b/1+a^2 \arctan y/x} e^{i\,b \phi/a}.
\end{equation}
This solution is multivalued, even for $b/a$ integer.  This type of variable
and solution could be pertinent in condensed matter applications, concerning
e.g. critical phenomena in multilayered films.

In general, our method provides us with ``local solutions'', not necessarily
defined for all of $\Bbb R^3$.  The solutions are not necessarily single valued
and they can have singularities for real values of the variable $\xi$.
Moreover, in view of the existence of the light cone it is sometimes necessary
to consider spacelike and timelike regions of space-time separately, since
solutions in these regions may differ.  Typical exemples of this phenomenon are
provided by algebras 17, 18 and 19.  We list two variables $\xi$ in
Table~\ref{tab1} for each of these, one valid for $t^2 - x^2 - y^2 > 0$, the
other for $x^2 + y^2 - t^2 > 0$.  In all cases we restrict to $a=0$, in order
to have $m=0$ in the table.  The simplest solutions are given by eq.~(4.17) for
$B<0$ and \eqref{eq5.3} for $B>0$.

In the case of algebra 17, the corresponding solutions are:
\begin{equation}
\begin{alignedat}{2}
&W = R_0 e^{\varepsilon b \sqrt{t^2-x^2-y^2}/(x+t)} e^{-i \,by/(x+t)},&\quad
&t^2 - x^2 - y^2 > 0,\\
&W = \tanh \frac{b}{2} \left( \frac{\sqrt{x^2+y^2-t^2}}{x+t} - \xi_0 \right)
e^{-i\,by/(x+t)},& &t^2 - x^2 - y^2 < 0.
\end{alignedat} \label{6.3c}
\end{equation}

The two solutions can be connected on the cone, however their derivatives will
be discontinuous in any case.

Similarly, for algebra 18 of Table~\ref{tab1} we find the elementary solutions
\begin{equation}
\begin{alignedat}{2}
&W = R_0 e^{b \arctan \sqrt{t^2-x^2-y^2}/y} e^{-i\, b \arctanh x/t},& \quad
&t^2 - x^2 - y^2 > 0,\\
&W = \tanh \frac{b}{2} \left[ \arctanh \,\frac{\sqrt{x^2+y^2-t^2}}{y} - \xi_0
\right] e^{-i\,b \arctanh y/t}& &x^2 + y^2 - t^2 > 0.
\end{alignedat} \label{6.4d}
\end{equation}

Finally, for algebra 19 of Table~\ref{tab1} we have
\begin{equation}
\begin{alignedat}{2}
&W = \tanh \frac{b}{2} \left(\arctan \frac{\sqrt{x^2+y^2-t^2}}{t} - \xi_0
\right)e^{-i\, b \arctan y/x},& \quad &x^2 + y^2 - t^2 > 0,\\
&W = W_0 e^{b/2 \arctanh \sqrt{t^2-x^2-y^2}/t} e^{-i\, b \arctan y/x},& &t^2 -
x^2 - y^2 > 0.
\end{alignedat} \label{6.5e}
\end{equation}

In many soliton-like problems in field theory we are
interested in solutions which are regular in $\Bbb R^3$, i.e.\ which
are valid at all times (though this condition
is sometimes relaxed a bit) and which are defined for
$-\infty<x<\infty$, $-\infty<y<\infty$. Among them particularly
important are those whose energy is finite (as they describe localised
``soliton-like" field structures).

If we restrict our attention to such field configurations we see that
we should consider the energy density for our fields.
As the energy density is given by
\begin{equation}
\rho={\vert W_t\vert^2+\vert W_x\vert^2+
\vert W_y\vert^2\over [1+\vert W\vert^2]^2},
\label{eq6.1}
\end{equation}
we see that this gives us
\begin{equation}
\rho={(\xi_t^2+\xi_x^2+\xi_y^2)\dot R^2\over [1+R^2]^2}
+{(\psi_t^2+\psi_x^2+\psi_y^2) R^2\over [1+R^2]^2},
\label{eq6.2}\end{equation}
where $\psi$ is given as in (\ref{eq3.1}) and
$\dot R={dR\over d\xi}$. We can rewrite
$(\psi_t^2+\psi_x^2+\psi_y^2)=(\alpha_t^2+\alpha_x^2+\alpha_y^2)
+\dot F^2(\xi_t^2+\xi_x^2+\xi_y^2)+2\dot F(\xi_t\alpha_t+\xi_x
\alpha_x+\xi_y\alpha_y)$ and then substitute the expression for $\dot F$
given by (\ref{eq3.9}) (when $m=0$).

To get the total energy we should integrate $\rho$ over all space
\begin{equation}
E=\int \rho\,dx\,dy.
\label{eq6.3}
\end{equation}
To perform this integration, in some cases, we can replace the integration
over $x$ and $y$ by an integration over $\xi$ and another conveniently
chosen variable (which may have a finite or an infinite range). Thus
in the cases of algebras 2, 3 and 19 of the Table~\ref{tab1} we can
use $\xi$ and an angle, while in the cases of
1, 4, 11, 12, 15 and 16 $\xi$ involves only $x$  and as
our variables of integration we can use $\xi$ and $y$.
Clearly in these latter cases the total energy of any nontrivial
solution is infinite.

The most extreme case corresponds to the algebra 10. In this case $\xi=t$,
energy
density is independent of $x$ and $y$ and so the total energy is infinite.
In this case $\phi^3$ of (\ref{eq1.3}) is given by $\phi^3={(1-R^2(\xi))/
(1+R^2(\xi))}$ and is independent of $x$ and $y$, while $\phi^1$
and $\phi^2$ depend on $x$ and $y$ only through $\alpha$. Thus
treating $\phi^{i}$ as components of a spin vector field (of unit
length) we see that this solution describes very coherent movements
of spins which move up and down in phase for all $x$ and $y$ and whose
movements in the horizontal plane are modulated by $\alpha$ and $F(t)$.

Similar spin wave interpretations can be given to other solutions.
In particular this is  the case when the symmetry variable is more complicated
than in the cases mentioned above. One can think of applications in
condensed matter physics, the theory of nematic liquid crystals etc.\
and even in cosmology. In some of such systems the orientation of
$\vec \phi$ does not matter; such cases can be described by a larger
class of our solutions. At the same time we can consider $W(x,y,t)$
as a Landau-Ginzburg field
which arises in many applications in condensed matter
physics (as can be checked the Landau-Ginzburg equation is very
similar to the equation derived from (\ref{eq1.4})). Indeed, at least one
version of the Landau Ginzburg equation has been treated using the group
theoretical techniques applied in this article.  The context was that of
magnetic phenomena in external fields.\cite{31}

However, returning to the field theory soliton-like applications,
in which case the reductions 2, 3, 17, 18 and 19 are particularly relevant, we
note
that using an angular variable of integration makes it more likely that
a given solution will describe a time evolution of a field configuration
of finite energy.

Clearly it would be desirable to analyse further the physical implications
of this and other solutions. We hope to be able to report on this in the near
furture.

\section*{Acknowledgements}

The research of A.~M.~G. and P.~W. was partly supported by research grants
from NSERC of Canada and FCAR du Qu{\'e}bec.

W.~J.~Z. wishes to thank the Nuffield Foundation and the British Council for
support of his visits to the Universit{\'e} de Montr{\'e}al
and the Centre de Recherches Math{\'e}matiques for its support and hospitality.

The authors thank W.~Hereman for helpful discussions.

\begin{landscape}
\begin{table}
\caption{\label{tab1}}
\squeezetable
\centering
\begin{tabular}{r*{10}{c}}
No & Algebra & $\xi$ & $\alpha$ & $p$ & $g$ & $h$ & $s$ & $l$ & $m$ &
$B$\\ \hline
1
& $K_1+aZ$, $P_2+bZ$
& $\sqrt{(t^2-x^2)}$
& $by- a\arctanh{x\over t}$
& $-{1\over \xi}$
& ${1\over \xi}$
& $0$
& $0$
& $-(b^2+{a^2\over \xi^2})$
& $0$ & $2(b^2 \xi^2+ a^2)$\\
2
& $P_0+aZ$, $L+bZ$
& $\sqrt{x^2+y^2}$
& $bt+a\arctan{y\over x}$
& $-{1\over \xi}$ & ${1\over \xi}$
& $0$
& $0$
& $-(b^2-{a^2\over \xi^2})$
& $0$ & $2(b^2 \xi^2- a^2)$\\
3
& $K_1+aZ$, $K_2+L$
& $\sqrt{t^2-x^2-y^2}$
& $-a\ln\vert x+t\vert$ & ${-2\over \xi}$
& ${1\over \xi^2}$
& $0$
& $0$
& $0$
& $0$
& $0$\\
4
& $K_2+aZ$, $P_0-P_2$
& $x$
& $-a\ln\vert y+t\vert$
& $0$
& $1$
& $0$
& $0$
& $0$
& $0$
& $0$\\
5
& $D+aZ$, $P_0-P_1$
& ${t+x\over y}$
& $a\ln\vert t+x\vert$
& ${-2\over \xi}$
& ${1\over \xi^2}$
& $0$
& $0$
& $0$
& $0$
& $0$\\
6
& $K_1+\epsilon P_2+aZ$, & ${y \ep}+\arcth{x\over t} +{1\over 2}\ln( t^2-x^2)$
& ${a \ep}y$
& $0$
& $1$
& $a$
& $0$
& $a^2$
& $0$
& $0$\\
& $P_0-P_1$
&
&
&
&
&
&
&
&
&
\\
7
& $D+\ep (K_2+L)+aZ$,
& ${2y\over t+x}+2\ep \ln\vert{t+x\over 2}\vert$
& $-a\ln\vert{{t+x}\over 2}\vert$
& $0$
& $1$
& $0$
& $0$
& $0$
& $0$
& $0$\\
& $P_0-P_1$
&
&
&
&
&
&
&
&
&
\\
8
& $D+aK_1+bZ$,
& ${x+t\over 2}y^{(a-1)}$, $a\ne 1$,
& ${1\over 2}b\ln(y)$
& ${2-a\over a-1}{1\over \xi}$
& $\xi^{(2-a)/(a-1)}$
& ${b\over 2(a-1)}{1\over \xi}$
& ${-b\over 2(a-1)^2}{1\over \xi^2}$
& ${b^2\over 4(a-1)^2}{1\over \xi^2}$
& $0$
& $0$\\
& $P_0-P_1$
&
&
&
&
&
&
&
&
&
\\
9
& $D+K_1+\ep(P_0+P_1)+aZ$,
& $e^{x+t} y^{-2\ep}$, \quad $\varepsilon = \pm 1$
& ${a\ep\over 2}(x+t)$
& $-{1+2\ep\over 2\ep}{1\over \xi}$
& $\xi^{-(1+2\ep)/2\ep}$
& $0$
& $0$
& $0$
& $0$
& $0$\\
& $P_0-P_1$
&
&
&
&
&
&
&
&
&
\\
10
& $P_1+aZ$; $P_2+bZ$
& $t$
& $by+ax$
& $0$
& $1$
& $0$
& $0$
& $-(b^2+a^2)$
& $0$
& $2(b^2+a^2)$\\
11
& $D+aZ$, $P_2$
& ${x\over t}$
& $a\ln t $
& $-{2\xi\over \xi^2-1}$
& ${1\over \xi^2-1}$
& $-{a\xi\over \xi^2-1}$
& $-{a\over \xi^2-1}$
& ${a^2\over \xi^2-1}$
& $0$
& $2a^2$\\
12
& $D+K_1+\ep(P_0+P_1)+aZ$, $P_2$
& $t+x -\ep \ln\vert t-x\vert$
& ${a\over 2}\ln\vert t-x\vert$
& $0$
& $1$
& $-{\ep a\over 4}$
& $0$
& $0$
& $0$
& ${a^2\over 8}$\\
13
& $D+aZ$, $P_0$
& $\arctan{y\over x}$ & $a\ln\sqrt{x^2+y^2}$
& $0$
& $1$
& $0$
& $0$
& $a^2$
& $0$
& $-2a^2$\\
14
& $D+aL+bZ$, $P_0$
& ${1\over 2}\ln({x^2+y^2}) -{1\over a}\arctan{y\over x}$
& ${b\over a}\arctan{y\over x}$
& $0$
& $1$
& $-{b\over 1+a^2}$
& $0$
& ${b^2\over 1+a^2}$ & $0$ & $-2{b^2a^2\over(1+a^2)^2}$\\
15
& $P_2+aZ$, $P_0+bZ$
& $x$
& $by+at$
& $0$
& $1$
& $0$
& $0$
& $(b^2-a^2)$
& $0$
& $2(a^2- b^2)$
\\ 16
& $D+aK_1+bZ$,  $P_2$,
& ${(t+x)^{a+1}\over (t-x)^{1-a}}$
& ${b\over 1+a}\ln\vert t-x\vert$
& $-{1\over \xi}$
& ${1\over \xi}$
& ${b\over 2\xi(a^2-1)}$
& $0$
& $0$
& $0$
& ${b^2 \over 2(a^2-1)^2}$ \\
&
& $a\ge0$, $a\ne1$,
&
&
&
&
&
&
&
& \\
17
& $D+aZ$, $K_2+L+bZ$
& $\frac{(t^2-x^2-y^2)^{1/2}}{x+t}$
& $-b{y\over t+x} +a\ln\vert t+x\vert$
& $0$
& $1$
& ${-a\over \xi}$
& $0$
& ${b^2}$
& $\frac{a}{\xi^2}$
& $2\left(\frac{a^2}{\xi^2} - b^2 \right)$\\
&
& $\frac{(x^2+y^2-t^2)^{1/2}}{x+t}$
& $-b{y\over t+x} +a\ln\vert t+x\vert$
& $0$
& $1$
& ${-a\over \xi}$
& $0$
& ${-b^2}$
& $\frac{a}{\xi^2}$
& $2\left(\frac{a^2}{\xi^2} + b^2 \right)$\\
18
& $D+aZ$, $K_1+bZ$
& $\arctan \frac{(t^2-x^2-y^2)^{1/2}}{y}$
& $-b\arctanh{x\over t}+\frac{a}{2} \vert t^2-x^2 \vert$
& $0$
& $1$
& $-\frac{a}{\tan \xi}$
& $0$
& $b^2 - a^2$
& $-\frac{a}{\sin^2 \xi}$
& $2\left(\frac{a^2}{\sin^2\xi} + b^2 \right)$\\
&
& $\arctanh \frac{(x^2+y^2-t^2)^{1/2}}{y}$
& $-b\arctanh{x\over t}+\frac{a}{2} \vert x^2-t^2 \vert$
& $0$
& $1$
& $-\frac{a}{\tanh \xi}$
& $0$
& $a^2 - b^2$
& $-\frac{a}{\sinh^2 \xi}$
& $2\left(\frac{a^2}{\sinh^2\xi} - b^2 \right)$\\
\\
19
& $D+aZ$, $L+bZ$
& $\arctan \frac{(x^2+y^2 - t^2)^{1/2}}{t}$
& $-b\arctan{y\over x}+ \frac{a}{2}\ln\sqrt{x^2+y^2}$
& $0$
& $1$
& $-{a\over\tan\xi}$
& $0$
& $-{a^2-b^2}$
& ${a\over \sin^2\xi}$
& $2\left(\frac{a^2}{\sin^2\xi} +b^2 \right)$\\
&
& $\arctanh \frac{(t^2-x^2 -y^2)^{1/2}}{t}$
& $-b\arctan{y\over x}+\frac{a}{2}\ln\sqrt{x^2+y^2}$
& $0$
& $1$
& $-{a\over\tanh\xi}$
& $0$
& ${a^2+b^2}$
& ${a\over \sinh^2\xi}$
& $2\left(\frac{a^2}{\sinh^2\xi} - b^2 \right)$\\
20
& $D-K_1+aZ$,
& ${\sqrt{x+t}\over y}$
& $at-{a\over 2}(x+t)+{b\over 2}\ln\vert x+t\vert $
& $-{2\over \xi}$
& ${1\over \xi^2}$
& $-{a\over 2\xi^3}$
& $0$
& $-{ab\over \xi^4}$
& ${a\over 2\xi^4}$
& $2({a^2\over 4\xi^2}+ab)$\\
& $P_0-P_1+bZ$
&
&
&
&
&
&
&
&
&
\\
21
& $D+aK_1+bZ$,
& ${1\over t^2-x^2-y^2} (t+x)^{2/(1-a)}$,
& ${b\over 1-a}\ln(x+t)$
& ${1\over 2}{3+a\over a+1}{1\over \xi}$
& $\xi^{-(3+a)/(2a+2)}$ & ${b\over a+1}{1\over 2\xi}$
& $0$
& $0$
& ${b(1-a)\over 4(a+1)^2}{1\over \xi^2}$
& $ {b^2\over 2(a+1)^2}\xi^{(1-a)/(1+a)}$\\
& $K_2+L$
& $a\ne\pm 1$,
&
&
&
&
&
&
&
&
\\
22
& $D-K_1+\ep(P_0-P_1)+aZ$,
& ${x^2+y^2-t^2\over 2(x+t)}-{\ep\over 2}\ln\vert x+y\vert$, & ${a\over
2}\ln\vert x+t\vert$
& $\ep$
& $e^{\ep\xi}$
& $-{\ep a\over 2}$
& $0$
& $0$
& ${a\over 2}$
& ${a^2 e^{-2\ep\xi}\over 2}$\\
& $K_2+L$
& $\varepsilon = \pm 1$
&
&
&
&
&
&
&
&
\\
23
& $D+{1\over 2}K_1+aZ$,& ${6(t-x)+(t+x)^3+6\ep y(t+x)\over [(t+x)^2+4\ep
y]^{3/2}}$
& $a\ln\vert (t+x)^2+4\ep y\vert$
& $\frac{5}{3}{-\xi\over 1-\xi^2}$
& ${1\over 1-\xi^2}^{5/6}$
& $\frac{2}{3}{a\xi\over 1-\xi^2}$
& $\frac{4}{9}{a\over 1-\xi^2}$
& $-\frac{4}{9}{a^2\over 1-\xi^2}$
& ${2a\over 9(1-\xi^2)^2}$
& ${8a^2\over 9}(1-\xi^2)^{-1/3}$\\
& $K_2+L+\ep(P_0+P_1)$
&
&
&
&
&
&
&
&
&
\end{tabular}
\end{table}
\end{landscape}

\begin{table}
\caption{\label{tab2}}
\centering
\begin{tabular}{*{9}{c}}
No & Algebra & $\xi$ & $\alpha$ & $\Delta\xi$ & $(\nabla\xi)^2$ &
$\Delta\alpha$ & $(\nabla\alpha,\nabla\xi)$ & $(\nabla\alpha)^2$\\
\hline 24 & $P_2+aZ$, $P_0-P_1+bZ$ & $x+t$ & $-bx+ay$ & $0$ & $0$ &
$0$ & $b$ & $-b^2-a^2$\\ 25 & $K_2+L+aZ$, $P_0-P_1+bZ$ & $x+t$ &
$bt-{2ay+by^2\over 2(x+t)}$ & $0$ & $0$ & ${b\over x+t}$ & $b$ &
$b^2-{a^2\over \xi^2}$\\ 26 & $D+K_1+bZ$, $K_2+L$ & $x+t$ &
$\frac{b}{2}\ln(t^2-x^2-y^2)$ & $0$ & $0$ & ${b\over t^2-x^2-y^2}$ &
${b(t+x)\over t^2-x^2-y^2}$ & ${b^2\over t^2-x^2-y^2}$\\ 27 &
$D-K_1+bZ$, $K_2+L$ & ${x+t\over t^2-x^2-y^2}$ & $\frac{b}{2}\ln(t+x)$
& $-2{(x+t)\over(t^2-x^2-y^2)^2}$ & $0$ & $0$ &
$-b{(t+x)\over(t^2-x^2-y^2)^2}$ &0\\ 28 & $D+K_1+bZ$, $P_2$ & $x+t$ &
$\frac{b}{2}\ln(t-x)$ & $0$ & $0$ & $0$ & ${b\over t-x}$ & $0$\\ 29 &
$D+K_1+bZ$, $P_0-P_1$ & $x+t$ & $\alpha=b\ln y$ & $0$ & $0$ & ${b\over
y^2}$ & $0$ & $-{b^2\over y^2}$
\end{tabular}
\end{table}

\end{document}